\documentclass[%
 reprint,
 amsmath,amssymb,
 aps,
]{revtex4-2}

\usepackage{graphicx}
\usepackage{dcolumn}
\usepackage{bm}

\usepackage{siunitx}
\DeclareSIUnit{\noisespectraldensity}{\volt^2\per\hertz}
\usepackage[hidelinks]{hyperref}
\newcommand{\orcid}[1]{\href{https://orcid.org/#1}{\includegraphics[width=10pt]{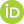}}}

\usepackage[utf8]{inputenc}

\begin{document}

\preprint{APS/123-QED}

\title{Injected power fluctuations for a non-equilibrium electronic disspative system in space}

\author{José A. Ogalde \orcid{0000-0001-9570-3187}}
 \email{jose.ogalde@ing.uchile.cl}

\author{Claudio B. Falcón \orcid{0000-0002-2730-1319}}

\author{Marcos A. Díaz \orcid{0000-0002-7701-5839}}
\affiliation{Faculty of Physical and Mathematical Sciences, University of Chile, Santiago 8370451, Chile. }

\date{\today}

\begin{abstract}
In this article we present an experimental study of the statistical properties for the injected power fluctuations of a dissipative system as a function of external environmental conditions. A Brownian motion analog is implemented using a series resistor and capacitor circuit with an Orstein-Ulhenbeck forcing. This system is tested in a controlled thermal bath at the laboratory, setting the bath temperature and different bath atmospheric pressures. The non-equilibrium system shows a higher correlation factor between the external forcing and the system response with increasing bath atmospheric pressure at constant temperature. These results were put to test in an uncontrolled bath such as space, by using a satellite orbiting at 505 km of altitude. A reduced version of the previous experiment was built to fit the satellite capabilities and was successfully integrated in the inner side of the satellite and then run in several locations of its orbit.
\end{abstract}


\maketitle

\section{Introduction \label{section:introduction}}

Out-of-equilibrium experiments are a rich and active subject of study in Physics. Out-of-equilibrium experiments have involved the use of a wide range of systems under observation such as colloidal particles in optical traps \cite{Gomez-Solano2010a,jop2008work,wang2005experimental}, mechanical harmonic oscillators \cite{joubaud2007fluctuation}, turbulent flows \cite{shang2005test,Falcon2008}, vibrating metallic plates \cite{cadot2008statistics, Garcia-Cid2015}, electrical circuits \cite{Falcon2009,Garnier2005,granger2015fluctuation,baiesi2016thermal}, DNA molecules \cite{frey2015reconstructing,mossa2009dynamic}, a gravitational wave detector \cite{bonaldi2009nonequilibrium} and granular media \cite{feitosa2004fluidized,kumar2011symmetry,mounier2012hatano,naert2012experimental, ciliberto2017experiments,Seifert2012a}. 
All of the previously mentioned experiments have measured the energy or power fluctuations of the system under out-of-equilibrium states and via its probability density function (PDF) with the goal of estimating system properties or testing theoretical findings \cite{ciliberto2017experiments,Seifert2012a}. 

Space is a natural hostile environment for electronics. The ultra high vacuum conditions make heat dissipation difficult via convection. This extreme situation, to which any satellite is subjected, is worsened by high doses of electromagnetic and/or particle radiation and cycles of significant temperature gradients. These conditions offer a natural laboratory for out-of-equilibrium experiments with platforms such as the Cubesat.

The Cubesat standard was conceived in 1999 as an educational tool mainly for universities \cite{mehrparvar2014cubesat}. Nevertheless, a few years after the first launches of Cubesats other organizations and companies were interested in the platform, noticing the opportunities that the platform might offer, with lower mission costs and  faster developing times \cite{poghosyan2017cubesat}. Initially these missions were mainly for technology demonstration. However, the development of the standard capabilities has currently allowed more scientific missions.    

Cubesats have also enabled new opportunities in space to countries with limited experience in space. The standard has had its impact also on latinamerican countries by developing missions and programs \cite{Diaz2016} based on this platform. In that regard, the Satellite of the University of CHile for Aerospace Investigation (SUCHAI-1) is the first Cubesat developed in Chile. Designed and built entirely by the Space and Planetary Exploration Laboratory (SPEL) at the University of Chile. The SUCHAI-1 Cubesat started as a seed project in 2011 with the hope to set the roads for a space program in the same university. SUCHAI-1 was placed in a polar sun-synchronous orbit at an altitude of 505 km orbit by a PSLV rocket, launched from Sriharikota, India in 2017 \cite{Gonzalez2018}.

The approach selected for this investigation is to replicate the experimental setup used in a study performed by \citet{Falcon2009}, where it was proven that a series resistor and capacitor circuit can be used as analog the Brownian motion to study fluctuations theorems for Ornstein-Uhlenbeck forcing by using a voltage source with pseudo-random noise. 
First this experiment was replicated and tested under controlled environmental conditions using a thermal vacuum chamber in the university laboratory. Following these tests, it was found a relation between the  injected power distribution of the system and the environmental atmospheric pressure that surrounds the system. Also, this test provided the necessary confidence to integrate this experiment into the Cubesat architecture in order to compare the injected power distribution between space environment and laboratory tests.

To our knowledge this kind of experiment has never been performed inside the Cubesat platform or space conditions. The use of passive electronics to study non-equilibrium fluctuations presents the advantage of being small compared with other parts for the satellite, in having low power consumption and more importantly it does not involve mechanical parts or chemical  compounds. All these qualities makes the experiment Cubesat-complaint \cite{mehrparvar2014cubesat}, low-risk and safe for the satellite, therefore being able to integrate it as payload under the SUCHAI's bus architecture explained in \citet{gonzalez2019architecture}.

This article is organized as follows: Section  \ref{section:experiment} explains how the circuit serves as a physical realization of a Langevin-type model for reader's convenience and provides the expected distribution for the power fluctuations.
Section  \ref{section:rcadhoc} presents the results of the tests performed in a thermal vacuum chamber to simulate space vacuum conditions for the circuit.
Section \ref{section:suchai-implmentation} remarks some important implementation details of the experiment to be integrated as a payload of the satellite. 
Section \ref{section:telemetry_section_label} shows the results obtained from the data downloaded from the satellite during its life-cycle and provides information of other sensors boarding the satellite such as temperature sensors and a particle counter. 
Finally Section \ref{section:conclusions} presents the main conclusions of this work.

\section{Brownian motion analog circuit \label{section:experiment}}
The experiment shown in this article was based on a previous work, in which an electronic circuit was used as a Langevin-type model driven with an Ornstein-Ulhenbeck forcing  to study the statistical properties of the injected power fluctuations under normal environmental conditions \cite{Falcon2009}. In this article we will put the former experiment under different atmospheric pressure and temperature to study their effects in the non-equilibrium distribution. For reader's convenience we will describe the principles of the experiment in this section. 

\begin{figure}
\centering
\includegraphics[width=0.95\linewidth]{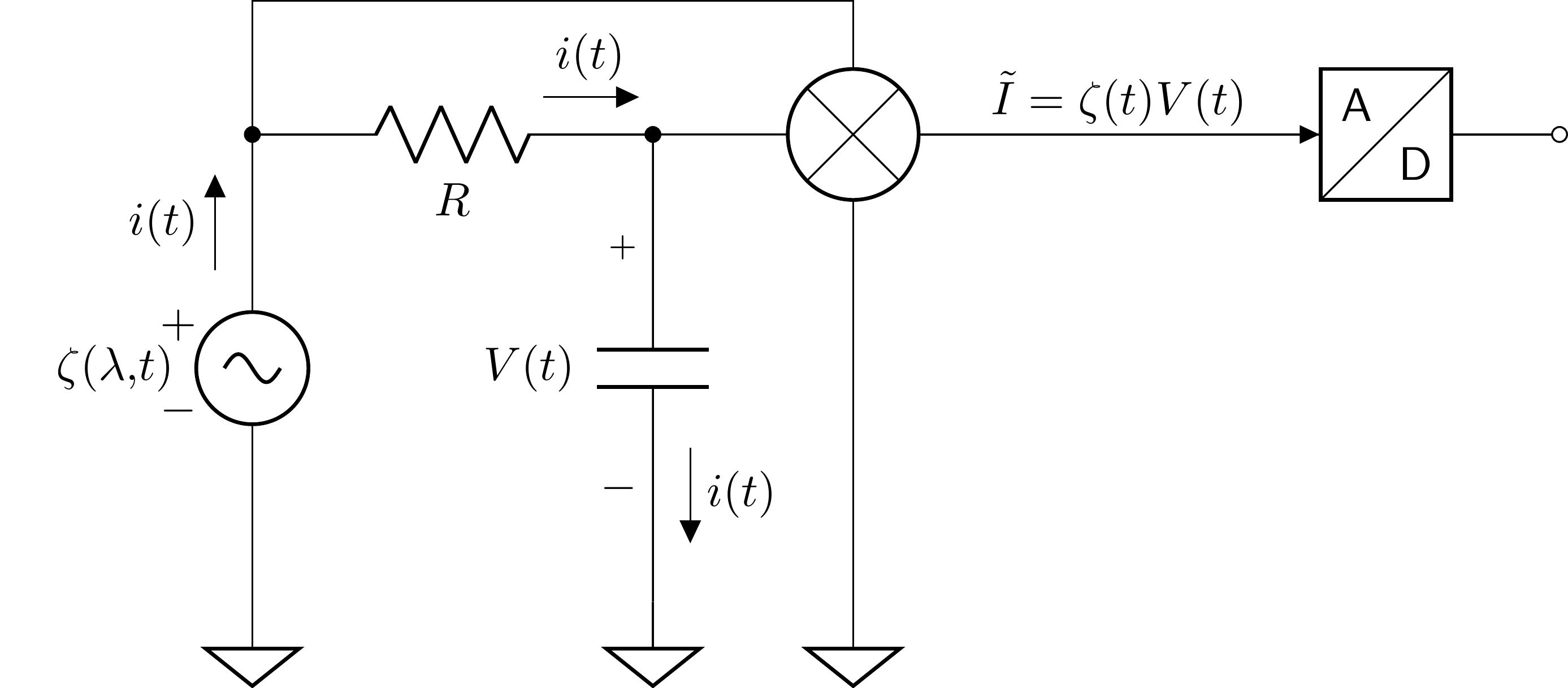}
\caption{Schematic of the experiment proposed by \citet{Falcon2009} replicated in this article.}
\label{fig:RC}
\end{figure}

The experiment proposed in the article of \citet{Falcon2009} consisted of a resistor R in series with a capacitor C driven to an out-of-equilibrium steady state with a pseudo random Gaussian noise. The zero mean Gaussian random noise $\zeta(\lambda,t)$ is generated with a spectrum analyzer and low-pass filtered at $\lambda = 5 $kHz. The damping ratio $\gamma=1/RC$ is a controlled variable, since C is fixed and R could be varied allowing the damping rate $\gamma$ to range between 50 Hz and 10 kHz. The voltage measured at the capacitor terminals  $V(t)$ is multiplied with the source noise voltage $\zeta (t)$ with an analog multiplier. The multiplied analog signal $V(t)\zeta(t)$ is proportional to the injected power and is acquired with an A/D acquisition card at 100 kHz sampling rate during 10 seconds with an oversampling factor of $\beta=10$. Figure \ref{fig:RC} shows the schematic of the circuit proposed in \citet{Falcon2009} and replicated in this article. 

The current in the resistor  $i(t)$ equals to the current in the capacitor leading to the well-known first order equation of a RC circuit. The noise $\zeta(\lambda,t)$ is the external force that drives the system and in circuital terms it can be treated as a voltage source. The noise has a Gaussian probability distribution, and can be modeled as zero-mean delta-correlated Gaussian white noise, which can be expressed with the following equations:
\begin{align}
	\langle \zeta(t) \rangle &= 0 \\
	\langle \zeta(t) \zeta(t') \rangle &= D \delta(t-t') \label{eq:cap1:delta-correlated-noise}
\end{align}

Where $D$ is the noise spectral density value and its measured in [\SI{}{\noisespectraldensity}]. Given that $\zeta (t)$ is a random process, then the capacitor voltage $V(t)$, current $i(t)$ and any other combination of these variables will also be a stochastic process.
Defining $\gamma=1/RC$, leads to the equation:
\begin{align}
\frac{dV(t)}{dt}  &= \gamma  \zeta(t) - \gamma  V(t) \label{eq:cap1:original-experiment-rc-series-circuit-model}
\end{align}

The equation (\ref{eq:cap1:original-experiment-rc-series-circuit-model}) is a realization of the Langevin equation used to describe the Brownian motion of a particle.  Multiplying equation (\ref{eq:cap1:original-experiment-rc-series-circuit-model}) with the capacitor voltage gives the energy balance equation (\ref{eq:cap1:original-experiment-rc-power-relation1}) and (\ref{eq:cap1:original-experiment-rc-power-relation2}):
\begin{align}
\frac{d}{dt}\Big(\frac{V(t)^{2}}{2}\Big) &=  \gamma \zeta(t)V(t)- \gamma V^{2}(t) \label{eq:cap1:original-experiment-rc-power-relation1} \\
\frac{d}{dt}\Big(\frac{V(t)^{2}}{2}\Big) &=   I(t) - L(t) \label{eq:cap1:original-experiment-rc-power-relation2}
\end{align}

In equations (\ref{eq:cap1:original-experiment-rc-power-relation1}) and (\ref{eq:cap1:original-experiment-rc-power-relation2}) the random process $I(t)=\gamma\zeta(t)V(t)$ is identified as the injected power which drives the energy budget of the system and $L(t)=\gamma V^{2}(t)$ is the dissipation of the system. The change of variable $\tilde{I}(t)= I(t)/\gamma = \zeta(t)V(t)$ is the unnormalized injected power, which corresponds to the multiplication of the forcing and the output of the circuit. 

\begin{figure}
\centering
\includegraphics[width=0.9\linewidth]{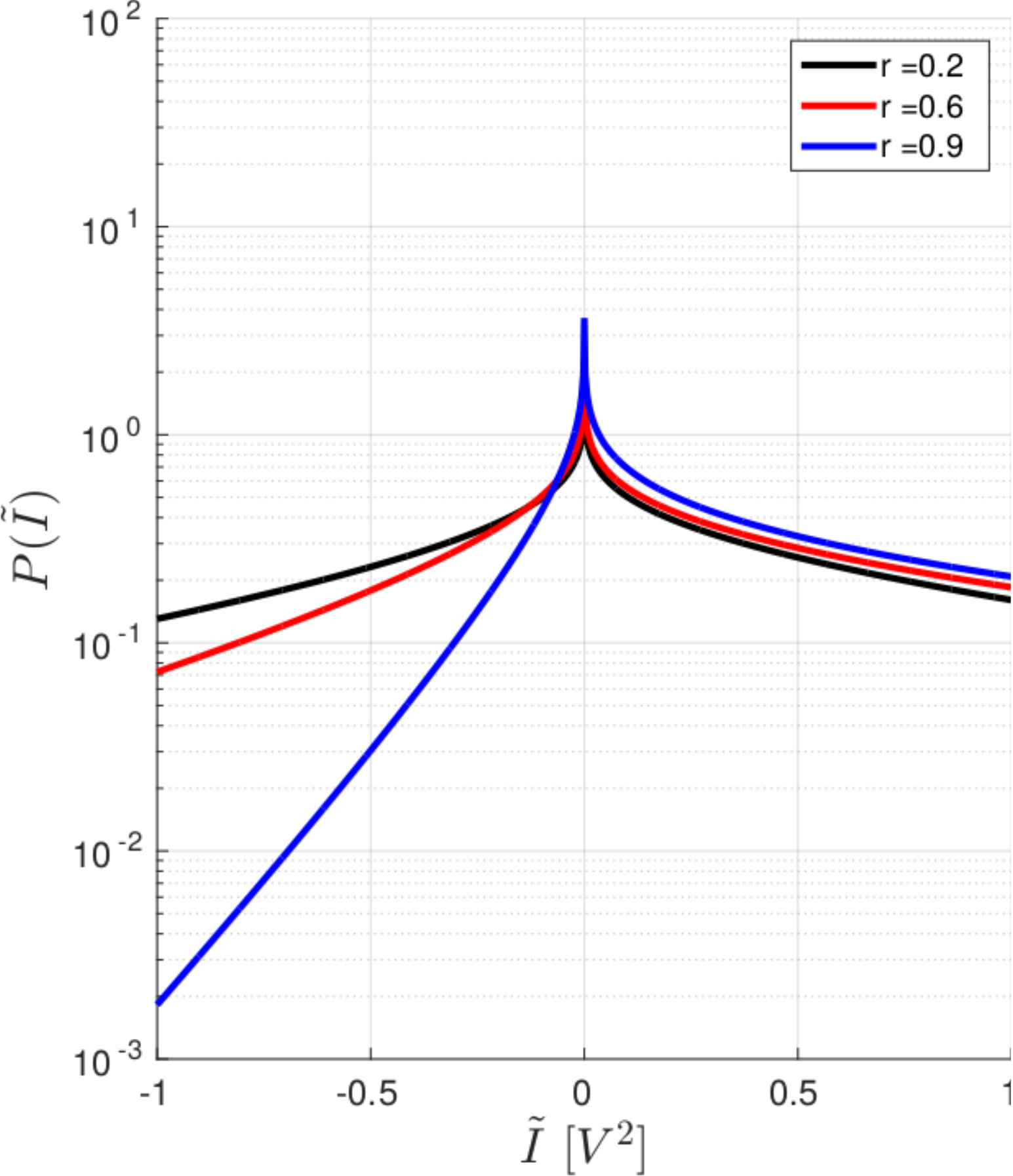}
\caption{Simulated probability distribution function for the unnormalized injected power $\tilde{I}$ of the out-of-equilibrium circuit at different values of correlation factor ($r$) between the input $\zeta(t)$ and output $V(t)$ voltages of the circuit (standard variations $\sigma_{\zeta} = 2$ V, $\sigma_{V} = 1 $ V  for this example).}

\label{fig:falcon2009-pdf0}
\end{figure}

It has been shown that the injected power gives information about the interaction of the system with its surroundings \cite{Falcon2008,Falcon2009,Farago2002,granger2015fluctuation}, since it provides the necessary energy to sustain the system in a nonequilibrium steady state. In this case, P($\zeta,V$) is the bivariate normal distribution of random variables $\zeta$ and $V$ with zero means. P($\tilde{I}=\zeta V$) is the distribution of the injected power with which is a function of the correlation factor $r=\frac{\langle \zeta V \rangle}{\sigma_{\zeta}\sigma_{V}}$ \cite{nadarajah2016distribution,Bandi2008} (see Eq. (\ref{eq:cap2:pdfIteorica})). The PDF is computed from the voltage measurements and is studied for each set of control parameters, which in  \citet{Falcon2009} are the damping rate $\gamma$ and the noise spectral amplitude D. For fixed values of D and the forcing bandwidth $\lambda$, the measurements show a direct relation between the circuital damping rate $\gamma$ and the correlation factor $r$ as given by $r=\sqrt{\frac{\gamma}{\gamma + \lambda}}$, therefore making the asymmetry of the PDF controllable by a combination of circuit parameters and noise spectra ($\gamma, D, \lambda$). The PDF of the injected power is described by Eq. (\ref{eq:cap2:pdfIteorica}) and shown in  Figure \ref{fig:falcon2009-pdf0} for a set of correlation factors $r$.

\begin{align}
P(\tilde{I}) &= \frac{1}{\sqrt{1-r^{2}}\sigma_{\zeta}\sigma_{V}} \text{exp} \Big[\frac{r \tilde{I}}{(1-r^{2})\sigma_{\zeta}\sigma_{V}}\Big] K_{0}\Big[\frac{|\tilde{I}|}{(1-r^{2})\sigma_{\zeta}\sigma_{V}}\Big] \label{eq:cap2:pdfIteorica}
\end{align}
Where $\tilde{I} = \zeta V$ is the unnormalized injected power, and $\sigma_{\zeta}$ and $\sigma_{V}$ are the standard deviations of the forcing ($\zeta$) and the capacitor voltage ($V$), respectively.

The former equation for correlation factor is only valid when environmental perturbations can be modeled as part of the white noise $\zeta$. In the following sections it will be shown that $r \neq \sqrt{\frac{\gamma}{\gamma + \lambda}}$ for very low atmospheric pressure, which means that microscopic interactions between the thermal bath and the system cannot be described using the former model of Eq. (\ref{eq:cap2:pdfIteorica}).

\section{Results from controlled environment \label{section:rcadhoc}} 

As previously mentioned in Section \ref{section:introduction}, a set of experiments using a thermal vacuum chamber were performed to simulate space conditions. The objective is to drive the circuit into a non-equilibrium steady state in the SUCHAI-1 satellite, but in a controlled environment. A thermal vacuum chamber allows testing devices under controlled temperature and pressure conditions, thus making it possible to simulate static space environments for the RC circuit.

\begin{figure}		
    \centering
	\includegraphics[width=0.55\linewidth]{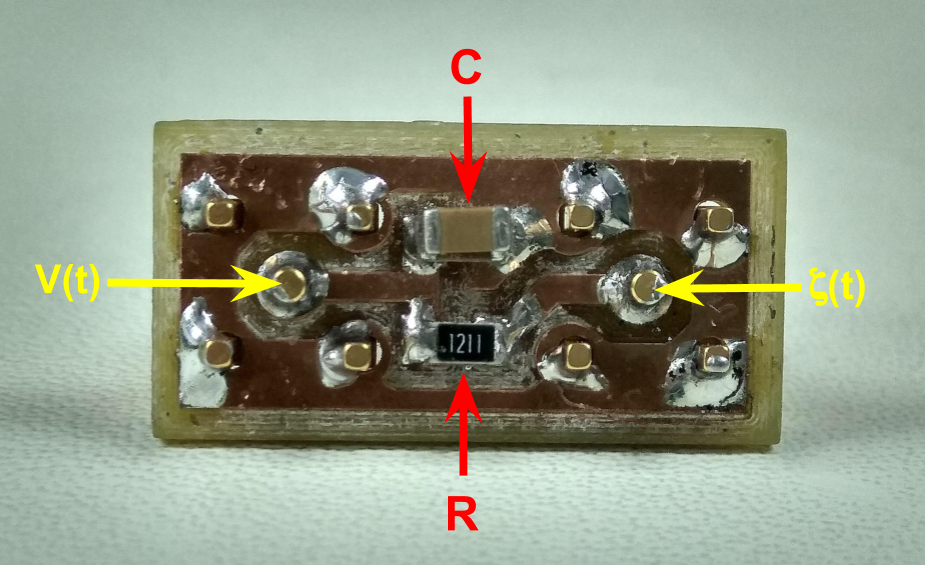}
	\caption{Photograph of the circuit tested inside the thermal vacuum chamber.}
	\label{fig:replicaRC-photo}
\end{figure}

The circuit was put in a thermal vacuum chamber to conduct these tests. We used the circuit shown in Figure \ref{fig:replicaRC-photo} with R = 1210 $\Omega$ and C =1.47 $\mu$F.

\begin{figure}[ht!]		
    \centering
	\includegraphics[width=0.9\linewidth]{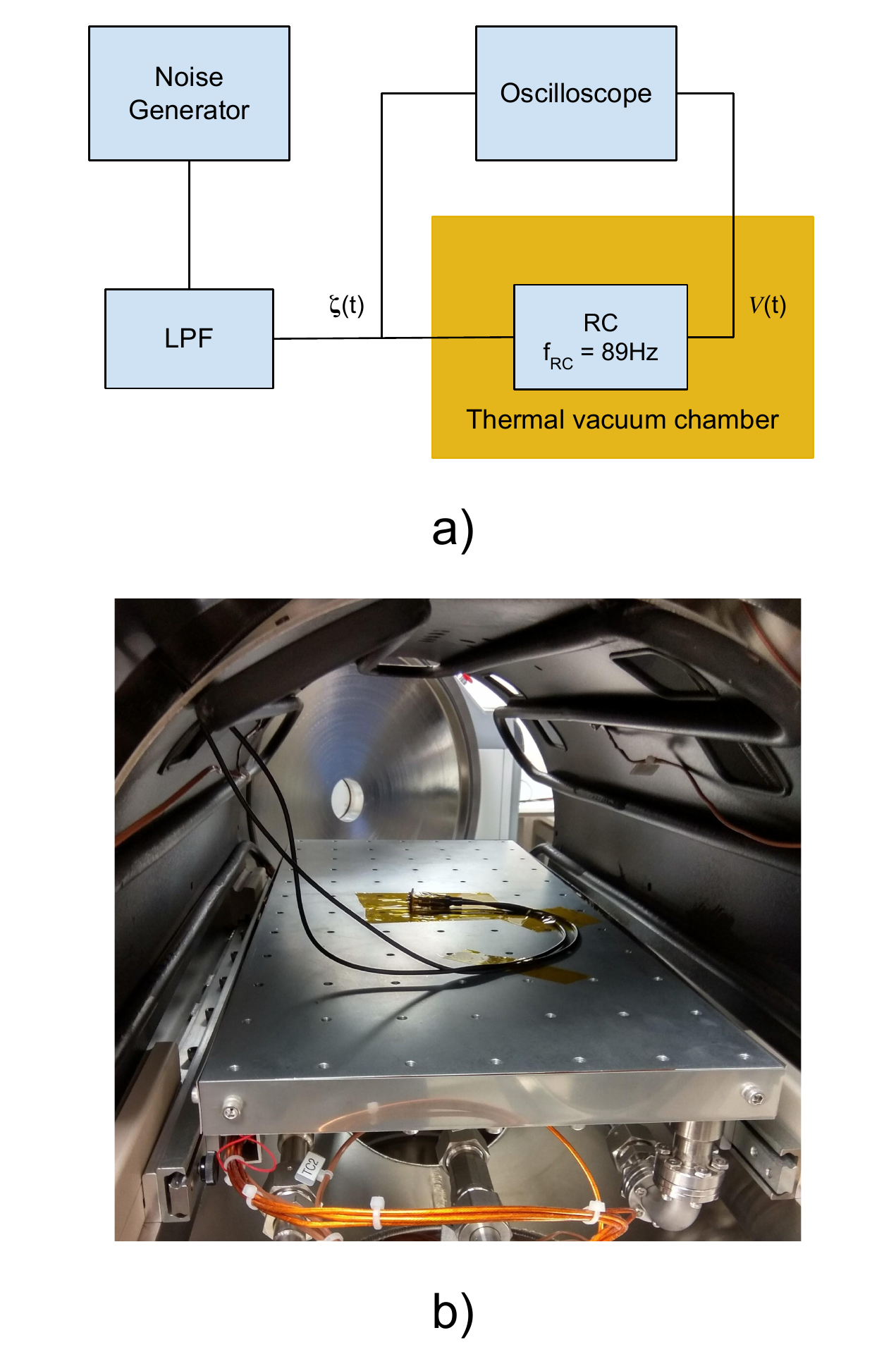}
	\caption{Tests in the thermal vacuum chamber a) Diagram of the experimental setup. b) Photograph of the RC circuit inside the thermal chamber. }
	\label{fig:thermo-vacuum-tests}
\end{figure}

The experimental setup is described in the system diagram shown in Figure \ref{fig:thermo-vacuum-tests}a. The circuit is driven into non-equilibrium steady state by a white noise generator using a standard laboratory signal generator. The voltages signals at the input of the circuit  $\zeta(t)$ and  at the capacitor terminals $V(t)$  are measured with an oscilloscope. These two instruments (signal generator and oscilloscope) are physically outside of the thermal vacuum chamber in order to control only the environment around the circuit. The forcing $\zeta(\lambda,t)$ is filtered at $\lambda = 1.8$ KHz  with a low pass filer. Dynamic range of the signal generator was tested from 16 V peak-to-peak down to 0.33 V. The voltages  were acquired at $F_{S} = 62.5$ MHz (or each $T_{S} = 16$ ns) for a total of $N_{S} = 125000$ samples.

For this experimental setup, two external environmental conditions are set within the thermal vacuum chamber configuration. Environment Condition 1 (EC1) sets the normal ground pressure conditions and Environmental Condition 2 (EC2) sets the space vacuum conditions while keeping the temperature constant in both scenarios.

\begin{itemize}
    \item Environmental Condition 1 (EC1): Pressure $P = 101.3 \ \text{kPa} = 760 \ \text{Torr}$ and temperature $T= \SI{22}{\celsius}$.
    \item Environmental Condition 2 (EC2): Pressure  $P =667 \mu\text{Pa} = 5 \mu\text{Torr}$ and temperature $T= \SI{22}{\celsius}$.
\end{itemize}

The random nature of the forcing $\zeta(t)$ can play an important role in the probability distribution of the injected power \cite{ciliberto2017experiments}. In several random driven experiments \cite{Farago2002,Farago2004,bellon2013measuring,Falcon2008,cadot2008statistics,Falcon2009,bonaldi2009nonequilibrium} the external random force acts itself as a thermal bath producing discrepancies to the steady state fluctuation theorems. As mentioned by \citet{bellon2013measuring} one is interested in the distribution of the power/work fluctuations done by the external random force which acts as part of the system, making it impossible to separate them as two uncorrelated systems. Our hypothesis is that a condition where the non-equilibrium steady state of the circuit is directly affected by the environment can be found \emph{if} the fluctuations of the system are of the same order of magnitude with respect to the changes in fluctuations induced by the interaction with the environment. One way to see this is by adjusting the dynamic range of the forcing $\zeta(t)$ until environmental effects are detected in the power distribution. For this purpose, the circuit was excited with a tunable white noise looking for the smallest dynamic range of the forcing where environmental perturbations and the forcing amplitude are in the same order of magnitude and therefore can be detected through the injected power distribution. We achieved this by running the experiment using different peak-to-peak voltages of the signal generator, which is equivalent to modify $D$ in Eq. (\ref{eq:cap1:delta-correlated-noise}).

\begin{figure}
	\centering
	\includegraphics[width=0.9\linewidth]{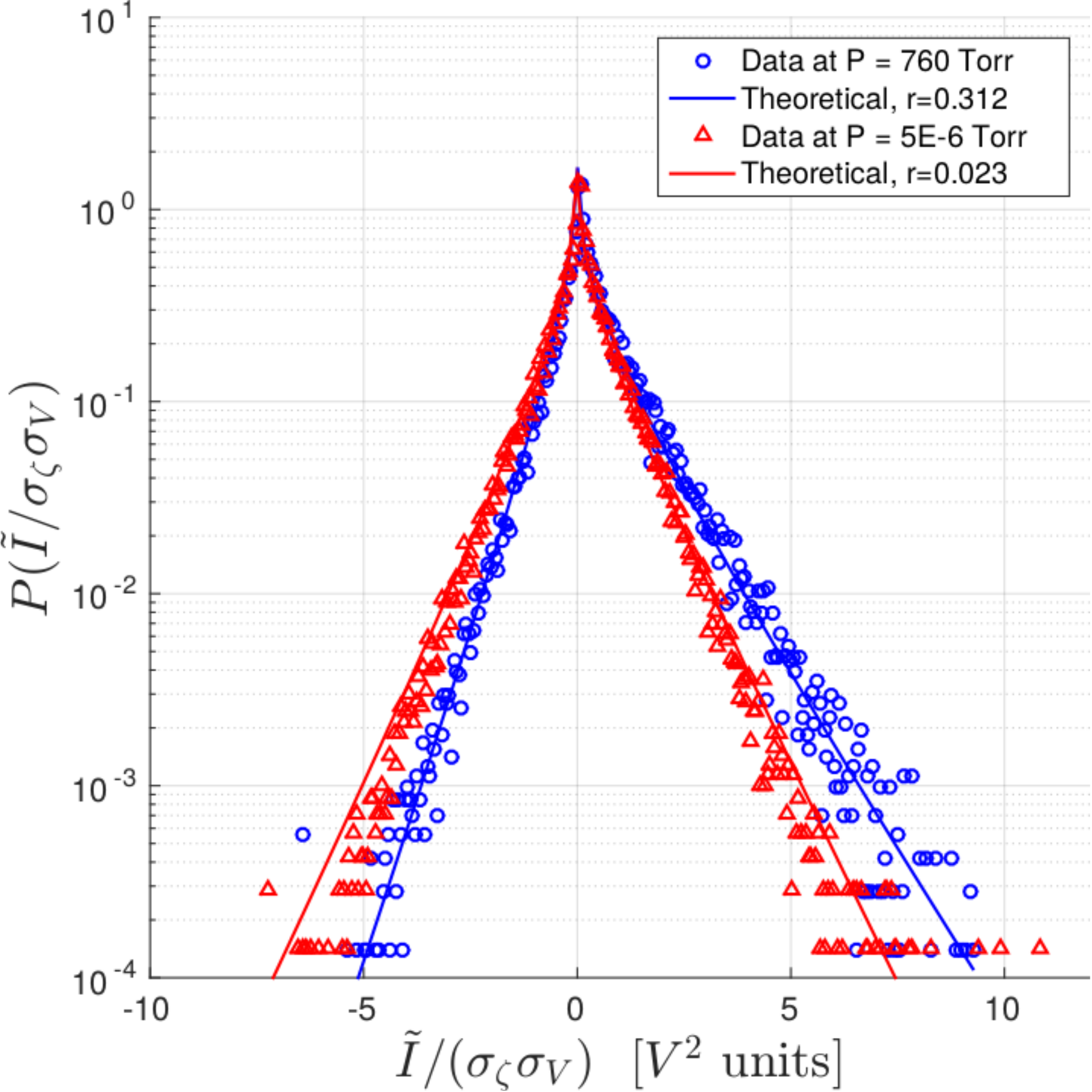}
	\caption{Probability density function of the normalized injected power of the new circuit for two different environmental pressures while keeping the parameters of the input noise (forcing) constant at 3.3 $V_{pp}$ for the dynamic range. Data with  blue \( \circ  \) markers is related to vacuum conditions ($5 \mu$ Torr) and data with red  $\triangle$ markers is related to normal pressure conditions (760 Torr /1 atm). Theoretical response given by Eq. (\ref{eq:cap2:pdfIteorica}) is superimposed (not fitted) on top of the data with a solid line with the same color.}	
	\label{fig:replica-falcon-lambda-1C8KHz1-6784e-11IDivBymeanI30-Aug-2018}
\end{figure}

With this setup the RC circuit maintains a non-equilibrium steady state while the oscilloscope records the signals and a computer is used to generate fluctuations statistics. The Figure \ref{fig:replica-falcon-lambda-1C8KHz1-6784e-11IDivBymeanI30-Aug-2018} shows the results of the two environmental conditions used with the thermal vacuum chamber. The distribution the power fluctuations are plotted both for normal ambient pressure (EC1: 760 Torr shown with a \( \circ  \) in blue) and for vacuum pressure (EC2: $5\cdot 10^{-6} \ \text{Torr}$ shown with a $\triangle$ in red). On top of these data, the theoretical response given by Eq. (\ref{eq:cap2:pdfIteorica}) is superimpossed (not fitted) with solid lines for the experimental correlation factor between $\zeta(t)$ and $V(t)$. 

Notice that the red curve at vacuum for EC2 has more negative fluctuations events ($\tilde{I}\leq 0$) when compared to the blue curve at normal atmospheric pressure EC1. Thus, showing that vacuum fluctuations have a more symmetrical distribution. This change in the distribution can be modeled as a change in the correlation factor $r$ of Eq. (\ref{eq:cap2:pdfIteorica}), where at sea-level pressure (blue curve) the correlation factor is greater than at vacuum (red curve). A physical explanation for this could be that the gas particles around the circuit provides a thermal bath helping to dissipate the energy provided by the external forcing $\zeta(t)$. Is important to note that this change in the injected power distribution was only found when the noise had a dynamic range less than 3.3 V peak-peak. For higher values, the power fluctuations did not reflect a change in the distribution of the injected power for different atmospheric pressures  of the chamber which proves our former hypothesis.

Also notice that the superimposed line in Figure \ref{fig:replica-falcon-lambda-1C8KHz1-6784e-11IDivBymeanI30-Aug-2018}, correspond to Eq. (\ref{eq:cap2:pdfIteorica}) for the PDF of the injected power as described in Section \ref{section:experiment}. Therefore, data in Figure \ref{fig:replica-falcon-lambda-1C8KHz1-6784e-11IDivBymeanI30-Aug-2018}  demonstrate that the Langevin model fits the experimental data in this two Environmental Conditions (EC1 and EC2) and that the changes in pressure can be modeled through different correlation factor values $r=\frac{\langle \zeta V \rangle}{\sigma_{\zeta}\sigma_{V}}$.

Given the reduced size of this experiment, an interesting potential application of this relation between the environment pressure with the correlation factor in the non-equilibrium steady state, would be for the small satellites like the femto-satellites \cite{izquierdo2011next}: satellites that have a total mass lower than 100 gr. One of the main ideas of the femto-satellites is to deploy several units from a satellite that is already in orbit and make a mesh of measurements for the atmosphere while they are falling. Hardware restrictions for these satellites makes it difficult to use standard pressure sensors, altimeters and GPS in the LEO orbit \cite{perez2016survey}, mainly due to the reduced size and energy budget. Femto-satellites meshes could provide pressure data of the atmosphere which can be applied to refine drag force in deorbiting models. The vacuum pressure tested for this experiment is 667 $\mu$Pa which equals to 200 km of altitude or more. That is a region that cannot be measured by traditional avionics altimeters, because barometric technology can only detect pressures near to 0.456 Pa which equals to 85 km of altitude \cite{grigorie2010aircrafts}, thus the use of non-equilibrium electronic systems could potentially offer a technological advantage to measure altitude in higher regions of the atmosphere for small-sized satellites. 

As a previous step, in this article we put this experiment into a Cubesat, which is also a small satellite with a total mass of 1kg. In particular this was carried out by SUCHAI-1 Cubesat of University of Chile as a payload.

\section{Adaptation of the experiment for satellite integration\label{section:suchai-implmentation}}

The experiment described in Section \ref{section:experiment} and \ref{section:rcadhoc} was integrated in SUCHAI-1 as a Cubesat payload and given the restricted capabilities of the Cubesat, the instruments occupied in section \ref{section:rcadhoc} to generate the random process $\zeta(t)$ and measure to signals had to be adapted to met the Cubesat restrictions. This sections describes those modifications which were necessary to integrate it into the Cubesat. 

\begin{figure}		
	\centering
	\includegraphics[width = 0.9\linewidth]{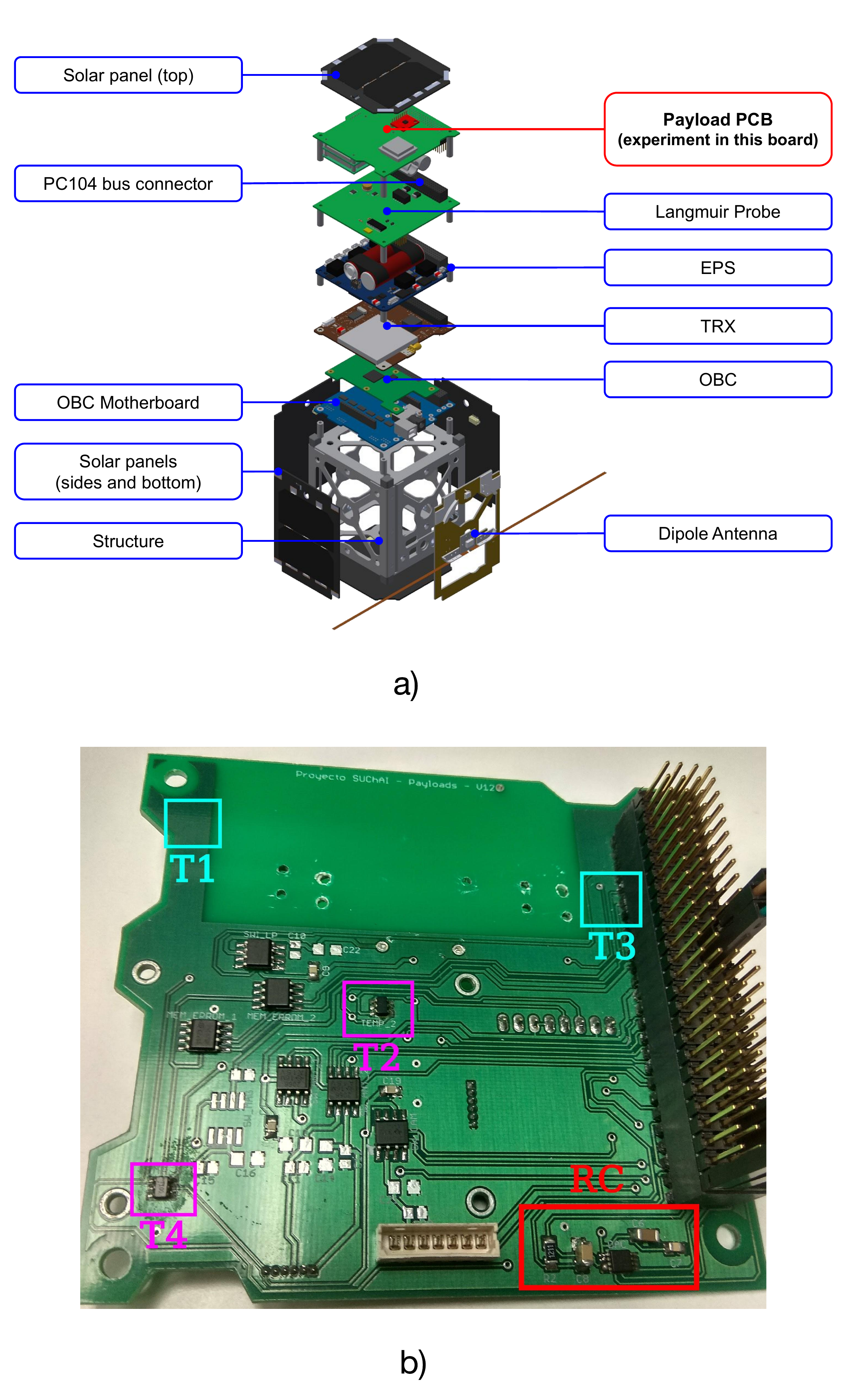}
	\caption{Implementation of the experiment as a Cubesat payload. a) Exploded view of SUCHAI-1 Cubesat sub-systems. b) Bottom-side photograph  of the board where the experiment is located: the resistor and capacitor of the experiment are labeled as \texttt{RC} (shown in red) and the temperature sensors are labeled as \texttt{T1, T2, T3, T4} (shown in cyan and magenta in top and bottom layers).}
	\label{fig:cap3:fig35}
\end{figure}

The main research focus of the SUCHAI-1 project is related to the measurement of the space environment and its impact on Cubesat components and systems \cite{Diaz2016,diaz2018preliminary}. 
Most of the SUCHAI-1 payloads were integrated in a custom-made board with all the payloads that are small in size like temperature sensors, a small camera, GPS module and memories. This experiment is also located in the same board. Figure \ref{fig:cap3:fig35}a shows a general exploded view of the satellite, where the main subsystems (OBC - On Board Computer, TRX - Transceiver, EPS - Electrical Power System, etc.) and Payloads are highlighted.

The resistor and capacitor are surface mount COTS components with measured and fixed values of $R = 1210$ [$\Omega$] and $C = 1.47 [\mu\text{F}]$. The voltage source occupied to implement the forcing $\zeta(t)$ was implemented by a pseudo random number generator in the integer domain of 16-bit plus a digital-to-analog converter which translates integers into physical voltages.

The dynamic range of the forcing $\zeta(t)$ was decided to have a constant value (D [\SI{}{\noisespectraldensity}] is constant) and the bandwidth $\lambda$ of the forcing is the controlled parameter. Since the circuit is a low pass filter, modifying the cutoff frequency of the forcing $\lambda$ is equivalent to controlling the amount of available energy to the circuit.  The method for setting a specific value of bandwidth $\lambda=\lambda_{0}$ is by changing the period register of the timer \texttt{T\_DAC}, which controls the time between two different values generated by the voltage source for $\zeta(t)$. Therefore, the noise bandwidth $\lambda$ is inversely proportional to the command argument $\texttt{arg\_value} \propto T_{\text{DAC}} \propto \lambda^{-1}$.

The energy consumption events of the injected power ($\tilde{I}>0$) will increase when the noise bandwidth get inside the pass-band of the circuit and the opposite will happen when the noise bandwidth is higher than the circuit cutoff frequency, thus making the shape of the injected power distribution controllable by this parameter, which is the same behaviour exposed in \cite{Falcon2009} and shown in the Figure \ref{fig:falcon2009-pdf0}, but by moving the $\lambda$ parameter instead of changing $\gamma$.

The sampling frequency of the satellite experiment was set to twice Nyquist rate ($F_{S} = 2(2\lambda)$ or  $\beta = 2$) A total of $N_s = 4000$ voltage samples per each run of the experiment equals to 8 KiB of data volume per run, which takes a complete SUCHAI-1 pass with line of sight to download (data link is of 9600 bps) \cite{Gonzalez2018}. Which means that the date of 1 run of the experiment can be downloaded in one day.

\section{Results from space environment and Discussion\label{section:telemetry_section_label}} 

SUCHAI-1 satellite was launched in a PSLV rocket from Sriharikota, India in 2017 which resulted in a circular polar sun-synchronous orbit at an altitude of 505  km. In this section we present the measurements and results obtained from the operations in SUCHAI-1 for the experiment presented in Section \ref{section:suchai-implmentation}.

\begin{figure}		
    \centering
	\includegraphics[width=0.9\linewidth]{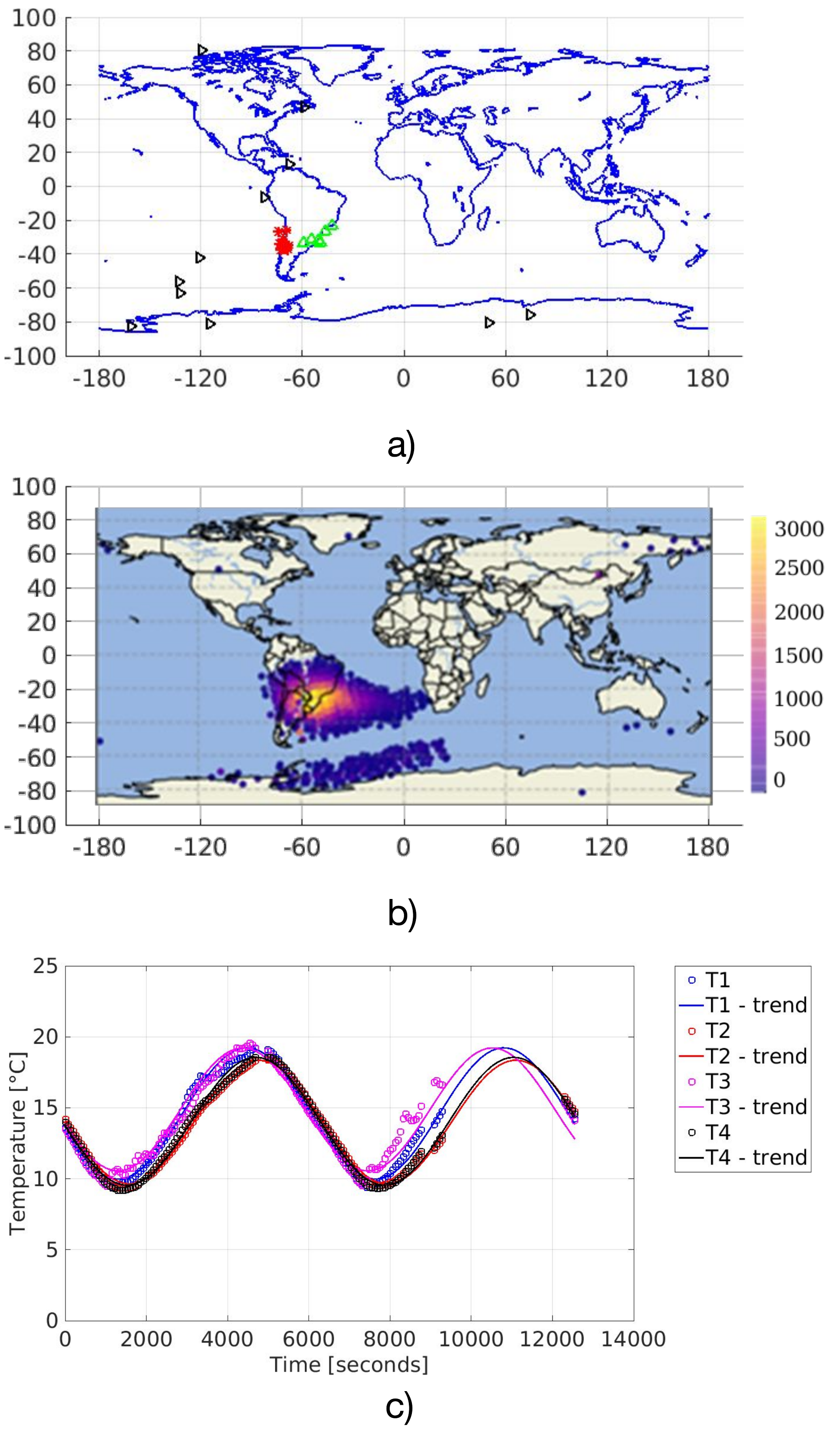}
    \caption{Experiment operations during Cubesat orbit a) Orbit locations for different executions of the experiment during Cubesat operations (Chile/GS are shown with * in red, SAA/Brazil are shown with * in green and other locations are with $\triangleright$ in black). b) Geomagnetic activity measured by the Cubesat as charged particles per minute (electrons and protons). c) Temperature cycles for the experiment during a typical orbit (four temperature sensors are inside the satellite \texttt{T1}, \texttt{T2}, \texttt{T3}, \texttt{T4} as described in Figure \ref{fig:cap3:fig35}).}
	\label{fig:cap4:mapa-operaciones-suchai}
\end{figure}

\begin{figure}
    \centering
    \includegraphics[width=0.9\linewidth]{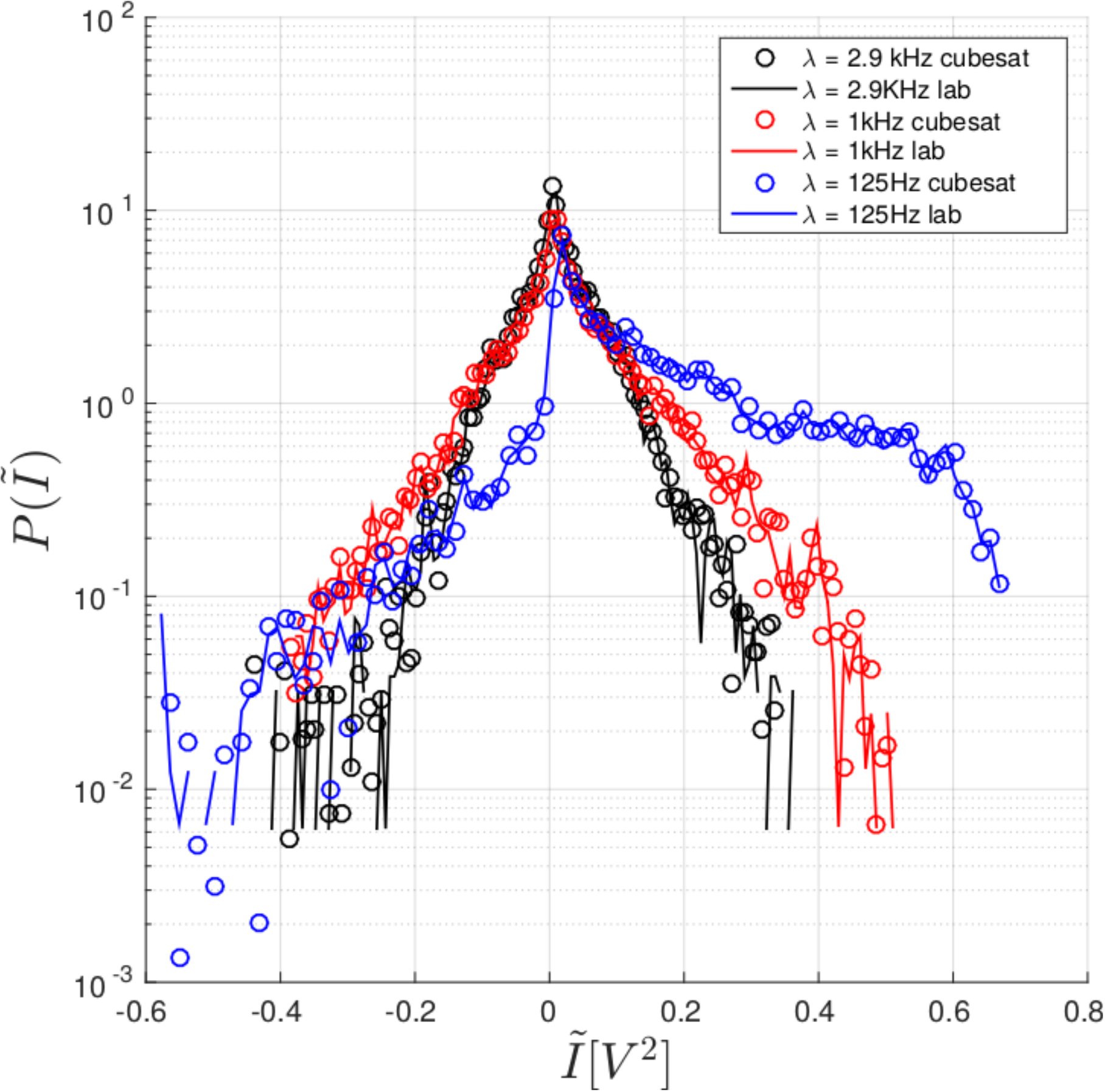}
    \caption{Probability density function of unnormalized injected power $\tilde{I}(t)$ for the circuit measured in the Cubesat orbit. The distribution for different forcing frequencies ($\lambda$) to the circuit are shown with a colorized \( \circ  \) marker. As a reference solid lines are added to represent the distribution obtained with a satellite replica in the laboratory.}
    \label{fig:results-histogram-downloaded-data}
\end{figure}

Figure \ref{fig:cap4:mapa-operaciones-suchai}a shows the location of the experiment executions. A total of 32 runs of the experiment were executed distributed between different earth locations: first above SUCHAI-1 ground station located in Santiago, Chile (shown with * in red), then at the center of the South Atlantic Anomaly (shown with green *) and finally at miscellaneous locations including the Pacific Ocean and Geomagnetic Poles (shown with $\triangleright$ in black). During SUCHAI-1 operations a wide sweep of 32 noise bandwidths were executed, starting from $\lambda = 14$ KHz to $\lambda = 15$ Hz and ended up focusing in a subset of three noise bandwidths $\lambda = 2.9 \text{KHz}, 1 \text{KHz}, 125 \text{Hz}$ for a total of 14 runs of this subset. The results of these three frequencies are summarized in Figure \ref{fig:results-histogram-downloaded-data}. 

In parallel to experiment executions, a particle counter was used in SUCHAI-1 to measure the Geomagnetic activity in the satellite orbit as shown in Figure  \ref{fig:cap4:mapa-operaciones-suchai}b. The particle counter measures  the amount of electrons and protons going through the satellite every minute. This  graph clearly shows the location of South Atlantic Anomaly (SAA), which has a peak in the southern part of Brazil. The graph was constructed by filtering out the regions that have less than 75 particles/min. Figure  \ref{fig:cap4:mapa-operaciones-suchai}b shows that locations where the experiment was executed had a peak of 3000 particles per minute when executed above the SAA and close to 100 particles/min when executed above the SUCHAI Ground Station in Chile. 

Temperature cycles during SUCHAI-1 orbit were also measured in parallel with four sensors \texttt{T1,T2,T3,T4} as described in Figure \ref{fig:cap3:fig35}b. These sensors reports temperature oscillations ranging from \SI{9}{\celsius} to \SI{19}{\celsius} with a peak-to-peak value of $\Delta T = \SI{10}{\celsius}$ measured at the inside of the Cubesat as shown in Figure \ref{fig:cap4:mapa-operaciones-suchai}c. This indicates that temperature conditions \emph{inside the Cubesat} are not out of COTS nominal temperature specifications which usually ranges between \SI{0}{\celsius} and \SI{70}{\celsius}.
 
The probability density function of the non-equilibrium steady states (non-equilibrium steady state) obtained each of the three bandwidths of the input noise $\zeta(\lambda, t)$ are shown in Figure \ref{fig:results-histogram-downloaded-data}: $\lambda = 2.9$ KHz (black \( \circ  \)), $\lambda = 1$ KHz (red \( \circ  \)) and $\lambda = 125$ Hz (blue \( \circ  \)) where the injected power fluctuations were measured in the locations shown in Figure \ref{fig:cap4:mapa-operaciones-suchai}a. The data of these executions is shown in Figure \ref{fig:results-histogram-downloaded-data} is complemented with the average distribution of the experimental runs obtained in the ground-based laboratory  prior to flight (solid lines), which allows to compare the response of the circuit in both environments.
Considering the cutoff frequency of the circuit is 89 Hz, the ratio between the noise bandwidth and the circuit cutoff frequency $(\lambda/f_{RC})$ ranges from 30 to 1.37 thus covering different filtering scenarios of the circuit.

The distribution in Figure \ref{fig:results-histogram-downloaded-data} shows asymmetrical exponential tails (non-Gaussian). This behaviour in injected power fluctuations has been predicted by non-equilibrium theory and have appeared in other out-of-equilibrium experiments \cite{Farago2002,Bandi2008,Gomez-Solano2010a,Falcon2008}, which is a clear indicator that a non-equilibrium steady state regime has been reached by the circuit. The curves resembles the PDF shown in Figure \ref{fig:falcon2009-pdf0} and also the results of \citet{Falcon2009}, where the PDF of the injected power is shifted towards the dissipation or positive events ($\tilde{I}>0$), by increasing the correlation factor $r=\frac{\langle \zeta V \rangle}{\sigma_{\zeta}\sigma_{V}}$ between the input/output signals. 

Notice that for $\lambda = 2.9$ KHz (black \( \circ  \)) there is a high level of symmetry between the positive events ($\tilde{I}>0$) and the negative events ($\tilde{I}<0$) of the injected power. As the noise bandwidth gets lowered to $\lambda = 1$ KHz (red \( \circ  \)) there is a very slight increase of positive events and thus a decrease in negative events, showing an increase in the level of asymmetry in the distribution. For a lower noise bandwidth $\lambda = 125$ Hz (blue \( \circ  \)) the number of positive events is dramatically increased showing a clear level of asymmetry between the positive and negative events. 
The curves of higher bandwidth $\lambda = 2.9$ KHz and $\lambda = 1$ KHz are more similar to the lower values of correlation factor ($r \leq 0.6$) curves shown in Figure \ref{fig:falcon2009-pdf0}.  For this experiment, the results shown in Figure \ref{fig:results-histogram-downloaded-data} demonstrates that the tails of the non-equilibrium steady state gets more asymmetrical, increasing power dissipation events as long as $\lambda$ approaches the circuit frequency. It could be observed that a range of values for the noise bandwidth $\lambda$ with respect to the circuit frequency $f_{RC}$ are useful when studying the non-equilibrium steady state displacement of the circuit. The same applies to demonstrate that noise bandwidth $\lambda$ can be used as an alternative to control the injected power distribution $\tilde{I}$ when the damping ratio $\gamma$ is fixed as in this circuit. 

By comparing the Cubesat data  and the laboratory data shown in Figure \ref{fig:results-histogram-downloaded-data}, it could be noted that there is a strong correlation between data obtained on ground and data downloaded as telemetry. Computing the average root mean squared error (RMSE) for each bin of the in-orbit data set with the laboratory data as reference, gives an average  bin error of $10.17 \%$, $6.0 \%$ and $4.21 \%$ for forcing signals of $\lambda = 2.9$ kHz, $1$ kHz and $125$ kHz respectively, considering laboratory probability distribution bins as reference.

This shows that there is no great variation between orbit and ground operations of the experiment, even when tested at different locations of the orbit with respect to ground laboratory executions. This low variance in the distribution of executions at different locations of the orbit compared with ground, gives us information about the reliability and stability of surface mount COTS electronics in space, specifically at the inside of the Cubesat. To verify this behaviour, other locations were tested as shown in Figure \ref{fig:cap4:mapa-operaciones-suchai}a with black marks, but no new information was obtained from executing the payload in those locations as expected.

These results demonstrate the capability of the Cubesat platform to drive a circuit into an out-of-equilibrium steady state by using COTS components and only using a fraction of the power and volume of a 1U Cubesat. Although the results obtained  from satellite operations are useful as a starting point for non-equilibrium experiments with Langevin-type model in space, it could not replicate the results obtained in the thermal vacuum chamber shown in Section \ref{section:rcadhoc}.
Hardware improvements in satellite instrumentation are needed in order to equal conditions with the results shown in Section \ref{section:rcadhoc}, which are still challenging for Cubesat technology. \emph{``What are the limits between a dissipative system and its environment for sustaining a  non-equilibrium steady state?"} is a non-trivial question that could be analyzed both theoretically and empirically. Simulation of non-equilibrium systems such as RLC networks and non-linear circuits like CMOS bistables, diodes or amplifiers in the out-of-equilibrium regime \cite{freitas2020stochastic} could be the following step for this kind of experiment put to test in hostile environments like space.

\section{Conclusion\label{section:conclusions}}
This work presents an experimental study of the statistical properties for the injected power for a dissipative system in two scenarios: a controlled thermal bath in a thermal vacuum chamber on ground and in a non-controlled environment in a Cubesat satellite at 505 km of altitude. This work also presents the first electronic out-of-equilibrium experiment performed in a Cubesat of this kind.

An empirical approach to study the environmental effects of the atmospheric pressure over the injected power distribution following a previous work on Langevin equation excited with an Orstein-Ulhenbeck forcing using an empirical series resistor and capacitor circuit to describe the fluctuation theorem's limitations.
The experiment revealed a relation between the atmospheric pressure of the bath that surrounds the system, showing more symmetric tails for the injected power at lower atmospheric pressure and asymmetrical tails for higher pressure, which can be mathematically explained by a higher correlation factor between the forcing and system response.

The satellite was successfully launched in 2017 to a polar sun synchronous low-earth orbit of an altitude of 505 km from ISRO facilities at Sriharikota, India. A correct  operation of the satellite  allowed 32 executions of the experiment in different locations of the planet including places close to: Chile, the South Atlantic Anomaly and the Geomagnetic poles. 
The ground-based experiment had to be adapted to Cubesat infrastructure in order to avoid changing the resistance and in this manner satisfying the restrictions of the Cubesat platform. The new implementation of the experiment was successfully integrated in the SUCHAI-1 Cubesat (1U) as one of its payloads.

Analysis of injected power fluctuations of the circuit has been generated from the experimental runs during the Cubesat orbit which shows that despite of the high Geomagnetic activity and ultra-low pressure conditions, the Cubesat was able to sustain a non-equilibrium steady state for the circuit during its orbit replicating the results of the experiment on which it was based. This demonstrates the Cubesat capability to provide the necessary infrastructure for electronic-based out-of-equilibrium experiments. 

Although the results previously obtained in a controlled environment on ground -using a thermal vacuum chamber- could not be replicated during satellite operations, the satellite discrepancy with ground tests opens the room for further research and developments to integrate experiments in space probes or satellites that could study space environment effects on electronics from the non-equilibrium physics or stochastic thermodynamics perspective. 

\begin{acknowledgments}
This material is based upon work supported by the Air Force Office of Scientific Research under
award number FA9550-18-1-0249. This work is also founded by grants CONICYT FONDECYT 1190005, CONICYT FONDEQUIP EQM150138 and CONICYT FONDECYT 1151476. The authors would like to thank the Faculty of Physical and Mathematical Sciences of the University of Chile for supporting the SUCHAI program. We also acknowledge the work and effort of the entire Space and Planetary Exploration Laboratory (SPEL) team in developing the SUCHAI missions. Special acknowledgments to Tomás Opazo for the support in the integration to the satellite and to Javier Rojas for the support in testing in the thermal-vacuum chamber.
\end{acknowledgments}

\end{document}